%% file: paper.tex
\documentclass[conference,compsoc]{IEEEtran}
\usepackage{pkgs}
\usepackage{svg}

\newif\ifprintcomments

\newcommand{\gtan}[1]{%
  \ifprintcomments
    \textcolor{blue}{{\em\bf GT:} #1}%
  \fi
}

\newcommand{\yongzhe}[1]{%
  \ifprintcomments
     \textcolor{orange}{{\em\bf YZ:} #1}%
  \fi
}

\newcommand{\trent}[1]{%
  \ifprintcomments
     \textcolor{purple}{{\em\bf TJ:} #1}%
  \fi
}

\def\BibTeX{{\rm B\kern-.05em{\sc i\kern-.025em b}\kern-.08em
    T\kern-.1667em\lower.7ex\hbox{E}\kern-.125emX}}

\newcommand{\para}[1]{\vspace{1ex} \noindent \textbf{#1}}
\newcommand{\vulName}{CIV}

\definecolor{lightbeige}{RGB}{245,245,220}

\newtcolorbox{insightbox}{
  colback=lightbeige,
  colframe=black,
  boxrule=0.5pt,
  arc=0pt,
  left=6pt,
  right=6pt,
  top=6pt,
  bottom=6pt,
  boxsep=0pt,
  sharp corners
}

\begin{document}


\title{SoK: Understanding the Attack Surface in Device Driver Isolation Frameworks}

\author{
Yongzhe Huang\textsuperscript{1}, 
Kaiming Huang\textsuperscript{1},
Matthew Ennis\textsuperscript{1},\\
Vikram Narayanan\textsuperscript{2},
Anton Burtsev\textsuperscript{2},
Trent Jaeger\textsuperscript{3},
Gang Tan\textsuperscript{1}\\
\\[2ex]
\textsuperscript{1}The Pennsylvania State University - \textit{\{yzh89, kzh529, mje5606, gtan\}}@psu.edu
\\[1ex]
\textsuperscript{2}University of Utah - \textit{\{vikram, aburstev\}}@cs.uath.edu
\\[1ex]
\textsuperscript{3}UC Riverside - \textit{\{trentj\}}@ucr.edu
}


\maketitle
\thispagestyle{plain}

\begin{abstract}
Device driver isolation is a promising approach for protecting the kernel from faulty or malicious drivers, but the actual security provided by such frameworks is often not well understood. Recent research has identified Compartment Interface Vulnerabilities (\vulName{}s) in userspace compartmentalized applications, yet their impact on driver isolation frameworks remains poorly understood. This paper provides a comprehensive survey of the design and security guarantees of existing driver isolation frameworks and systemizes existing \vulName{} classifications, evaluating them under driver isolation. The analysis shows that different classes of \vulName{}s are prevalent across the studied drivers under a baseline threat model, with large drivers having more than 100 instances of different \vulName{}s and an average of 33 instances across the studied drivers. Enforcing extra security properties, such as CFI, can reduce the number of \vulName{}s to around 28 instances on average. This study provides insights for understanding existing driver isolation security and the prevalence of \vulName{}s in the driver isolation context, and extracts useful insights that can provide security guidance for future driver isolation systems.
\end{abstract}


\input{tex/intro}

\input{tex/related_works}
\input{tex/background}

\input{tex/threat_model}

\input{tex/meth}
\input{tex/eval}

\input{tex/conclusions}
\input{tex/acks}

\bibliographystyle{plain}


\input{tex/appendix}
\vspace{12pt}
\end{document}

%% file: tex/intro.tex
\section{Introduction}
\label{sec:intro}
\vspace{-0.5em}
Today's Linux 6.5 kernel contains roughly 9,921 device drivers, accounting for 70\% of its source code~\cite{kerneldrvrs}, nearly doubling since 2013. Developed by third parties who often lack an incomplete understanding of the kernel's programming and security idioms, device drivers are a primary source of defects in modern OS kernels~\cite{chen+:apsys11}. With over 80,000 commits per year, the fast-evolving device driver codebase represents the largest attack surface in modern kernels, making up \textbf{16--59\%} of all reported Linux kernel vulnerabilities since 2014 (\autoref{tab:driver-cves}).

To prevent device driver flaws from impacting kernel security, \textit{driver isolation} has been proposed, with a history dating back to the 1977 Multics report~\cite{multics-report}. This technique runs drivers in an isolated environment, confining faults and preventing single-point failures. While easily achieved through clean-slate kernel redesigns~\cite{l3, minix, seL4, nexus}, efforts have focused on isolating drivers from monolithic kernels~\cite{nooks, xfi, lxfi, lvds, lxds, side}. However, many attempts have failed due to the \textit{performance} and the \textit{complexity} of breaking apart the kernel's monolithic, shared-memory code~\cite{nooks, oskit, lxds, lvds, lxfi}. Recently, the balance has started to change with efficient hardware isolation mechanisms, such as memory tagging~\cite{sfitag} and VMFunc~\cite{intel-vmfunc-patent}, and scalable automated decomposition approaches, such as static analyses for automating kernel-driver code and data separation~\cite{ksplit, microdrivers}.

\begin{table}[H]
\centering
\footnotesize
\begin{tabular}{lccc}
\toprule
\textbf{Year} & \textbf{Total CVEs} & \textbf{Driver CVEs} & \textbf{Percentage} \\ 
\midrule
\textbf{2014} & 130 & 22  & \textbf{16.9\%} \\
\textbf{2015} & 80  & 15  & \textbf{18.8\%} \\
\textbf{2016} & 216 & 65  & \textbf{30.1\%} \\
\textbf{2017} & 452 & 267 & \textbf{59.1\%} \\
\textbf{2018} & 180 & 44  & \textbf{24.4\%} \\
\textbf{2019} & 291 & 154 & \textbf{52.9\%} \\
\textbf{2020} & 129 & 38  & \textbf{29.5\%} \\
\textbf{2021} & 162 & 42  & \textbf{25.9\%} \\
\textbf{2022} & 310 & 92  & \textbf{29.7\%} \\
\textbf{2023} & 263 & 75  & \textbf{28.5\%} \\
\bottomrule
\end{tabular}
\vspace{-0.75em}
\caption{Linux kernel CVEs versus those in its device drivers.}
\label{tab:driver-cves}
\end{table}

Despite recent advances, the security benefits of driver isolation remain unclear. Isolation frameworks limit driver privileges, but restrictions vary based on threat models. For example, Nooks~\cite{nooks} assumes buggy but not malicious drivers, allowing isolated drivers to read kernel pages and run privileged instructions, which cannot confine attacks from malicious or compromised drivers. Understanding the threat models and isolation guarantees of different frameworks is therefore crucial.

It is also important to understand the attacks that are still possible after isolation. Malicious drivers may still compromise kernel compartments by exploiting the isolation interface, as demonstrated by Lefeuvre et al.~\cite{assess_civ} in compartmentalized user-space applications. These Compartment Interface Vulnerabilities (\vulName{}s) occur because compartmentalized components may still share/communicate data with a  domain that can be used to launch attacks. Chien et al.~\cite{CIVScope} identify a specific type of \vulName{}s for the kernel-driver interface, showing that malicious isolated drivers can supply corrupted data to kernel memory API calls, compromising the confidentiality and integrity of kernel data. Other researchers have identified additional security issues~\cite{assess_civ,RLBox, protecting_comm_kernel_from_vulnerable_driver, CIVScope, evolve_OS_driver_secure_interface}, suggesting more types of \vulName{}s may impact the kernel-driver interface. A comprehensive study of \vulName{} types and their impact on kernel security is needed.

\vulName{}s pose significant threats to the kernel despite driver isolation, but the effectiveness of attacks using \vulName{}s varies depending on the defenses applied in driver isolation. For instance, a \vulName{} enabling buffer index corruption may succeed in frameworks lacking index validation, but can be mitigated by those implementing shared data validation~\cite{protection_info_computer_system}.

Given the variety of types of \vulName{}s, evaluating them under different threat models assuming diverse isolation protections is crucial. Previous studies, such as Conffuzz~\cite{assess_civ}, estimated \vulName{} prevalence based on a basic isolation model with limited security properties, such as memory safety and control flow integrity. {Some driver isolation projects, however, have multiple kinds of protections to enforce security properties on the isolation interface. As a result,} it is important to evaluate attack viability against more robust driver isolation systems with enhanced security measures.

In this study we aim to answer the following questions: (1) What are the security properties enforced by existing driver isolation frameworks? (2) What are the types and prevalence of \vulName{}s at the kernel-driver isolation interface? (3) How effective are the existing driver isolation frameworks against the various types of \vulName{}s? (4) How do security properties enforced by software hardening techniques, such as  memory safety or CFI, reduce the threat of \vulName{}s? 

To answer these questions, our work systemizes the knowledge of the attack surface and defenses at the kernel-driver interface and makes the following contributions:

\begin{itemize}
\item We systemize the threat models and  security properties of existing driver
  isolation frameworks. 

\item We summarize and augment existing CIV classifications \cite{CIVScope,
    assess_civ, RLBox} with general and new driver-isolation-specific CIV classes. We describe   these new classes with concrete examples.

\item We develop static analyses to classify and quantify \vulName{}s for a set of representative drivers.

\item  We evaluate \vulName{}s under a baseline threat model that only assumes basic driver defenses and security properties.  And we discuss how enforcing additional security properties 
may help in mitigating \vulName{}s. 
\end{itemize}

\para{Target interface} We focus on classifying \vulName{}s at the kernel/driver isolation interface. However, the threat of CIVs goes beyond and applies to any compartmentalized software, including isolating other kernel components (e.g., the file subsystem) and isolation in user-space programs. 
Our study focuses on device driver isolation for several reasons. (1)  Kernel-driver interfaces are
the de-facto isolation boundary by the majority of previous kernel isolation frameworks due to the high number of vulnerabilities and required privileges in drivers. (2) Device drivers normally have clear compartment interfaces to communicate with the kernel, making analysis tasks easier, as we will discuss in detail later in Section~\ref{sec:drv_isolation_overview}. (3) The driver and kernel interfaces normally pass large and complex objects, typically containing heterogeneous types of data, making observations extracted by studying device driver compartmentalization valuable for compartmentalization in other domains with less complex interface data.  While we leave the study of CIVs on other interfaces to future work, we believe that the conclusions we draw in this study are generalizable to other kernel and user-space interfaces.

%% file: tex/related_works.tex
\section{Related Systemization} \label{sec:related_works}
\vspace{-0.5em}
Shu et al.~\cite{shu_isolation_security_study} provided a systematization of general isolation techniques. Their work established a taxonomy of isolation mechanisms and policies, but did not classify nor quantify interface vulnerabilities, the focus of this study. Lim et al.'s~\cite{lim2024_kernel_compart_sok} provided a recent systematization of kernel compartmentalization techniques. They focus broadly on securing monolithic kernels through various compartmentalization strategies, including a deep coverage of emerging hardware features that can benefit future driver isolation approaches. However, they do not focus on interface vulnerabilities. Our work in comparison provides taxonomies for understanding interface vulnerabilities, for which we provide static analysis methods for  evaluation and quantification in the driver context.

Lefeuvre et al.~\cite{lefeuvre2024sok} conducted a systematization of software compartmentalization, examining existing compartmentalization on policy definition methods, compartmentalization abstractions, and enforcement mechanisms. Among many observations, their high-level observation on the importance of interface vulnerabilities aligns with ours. In comparison, by focusing on the interface vulnerabilities of kernel-driver interfaces, we are able to devise static analyses to quantify different types of interface vulnerabilities, making the threat concrete. In addition, we systemize a set of security properties enforced by driver isolation and able to perform quantification under different assumptions of security properties.



%% file: tex/background.tex
\section{Driver Isolation Overview}
\label{sec:drv_isolation_overview}
\vspace{-0.3em}
In this section, we review existing driver isolation frameworks, classifying them based on their core isolation techniques. Aligning with the scope of this paper, we focus solely on the work specific to driver isolation in monolithic operating systems. Consequently, designs involving different kernel architectures, such as microkernels\cite{l3, seL4, minix, sawmill}, are outside the scope of this survey.  Additionally, we do not discuss systems that focus on isolating general kernel modules\cite{hacks, iski, virtuos}. 
Overall, we have identified 20 most impactful projects in the area of driver isolation published in major OS and security venues over the past two decades. We summarize these systems in Table~\ref{tab:drv_iso_frameworks_summary}. 

\subsection{Sandbox architecture}
\label{subsec:sanbox_architecture}
\vspace{-0.5em}
All the driver isolation frameworks assume a \textit{sandbox}~\cite{assess_civ} architecture, in which the kernel is trusted and the driver is not. This is opposite to the \textit{safebox} architecture~\cite{assess_civ, iski, skee, kcofi, iago}, which instead assumes the isolated component is trusted and should be protected from the rest of the system. 

\subsection{Driver threat models}
\label{subsec:drv_threat_model}
\vspace{-0.5em}
All the driver-isolation projects we have surveyed assume two types of threat models:
\begin{enumerate}
    \item \textbf{Buggy}: Many driver isolation frameworks~\cite{nooks, unmodified-device-drivers, SafeDrive, microdrivers ,decaf, lxds, lvds} assume that a driver may contain bugs that can cause the entire kernel to malfunction or hang. This threat model primarily focuses on the availability of kernel services, rather than considering an active attack scenario. The goal of isolation in this context is to prevent a buggy driver from crashing or hanging the kernel, ensuring that other kernel services remain available.
    \item \textbf{Exploitable}: In this threat model, a device driver is assumed to contain bugs that can be exploited by an attacker to launch attacks on the kernel~\cite{xfi, lxfi, sud, huko, protecting_comm_kernel_from_vulnerable_driver, bgi, driverjar, sfitag}. These bugs often allow the attacker to gain powerful read or write primitives, which can be used to compromise the confidentiality or integrity of the kernel data. Isolation mechanisms aim to confine the impact of such exploits to the isolated driver, preventing the attacker from compromising the  kernel.
\end{enumerate}

\subsection{ Driver isolation boundaries}
\label{subsec:iso_boundary}
\vspace{-0.5em}
Traditionally, the device driver has two natural boundaries. The first is for \textit{driver/kernel} interaction, and the other is for \textit{driver/device} interaction. The kernel/driver boundary is formed by a set of kernel functions imported by the driver, including kernel and device libraries, kernel services, etc.~\cite{understanding_modern_device_driver}, and a set of functions the driver exports to the kernel to extend the kernel's functionality via a set of function pointers and the \lstinline{EXPORT_SYMBOL} macro. This natural boundary is used as the de facto isolation boundary between the kernel and a driver, with a few exceptions where only a subset of the driver code is isolated~\cite{microdrivers, decaf, twindrivers}. The systems that choose a different isolation boundary do so to address limitations of traditional isolation, such as performance. For example, Microdrivers~\cite{microdrivers}, decaf~\cite{decaf}, and Twindrivers~\cite{twindrivers} leave performance critical driver code inside the kernel to reduce domain switching overhead, achieving better performance. 

The driver/device boundary is defined by operations that interact with the device hardware, including access to memory-mapped I/O (MMIO) regions, I/O ports (on x86 architectures), Direct Memory Access (DMA), etc.~\cite{understanding_modern_device_driver}. By abusing the interaction with the device, a malicious driver could potentially exploit the driver/device interface to circumvent the isolation. For example, abusing DMA may allow the driver to access any kernel memory through the device. To mitigate this risk, some isolation projects~\cite{sud, lxds, lvds} 
prevent devices from accessing arbitrary physical memory via DMA using the IOMMU. This approach sets up page tables in the PCI device access module for each PCI device, effectively confining the range of addresses accessible to each device. 
While studying \vulName{}s for the driver/device interface is also important, we observe that the device states that can be corrupted by the driver are highly device-dependent. For instance, each device has its own set of registers and configuration settings that control its behavior, and the set of corruptible registers varies across devices. Therefore we leave the study of this interface to future work.

\subsection{Interface data} \label{subsec:iso_drv_shared_data}
\vspace{-0.5em}
Even under isolation, drivers and kernels still communicate by sharing data through the kernel/driver interface, typically via the arguments to and return values from invocations to interface functions or through global variables. However, not all data that could be referenced by the pointers passed across the interface needs to be shared for correct execution. For instance, when a reference to a complex structure is passed through an interface function, it may be the case that only a small subset of the structure's fields are actually needed. If a driver isolation framework imposes no restriction on the data synchronization, it may lead to the problem of \textit{oversharing}. Various driver isolation projects propose solutions to address this issue. Projects based on object copying~\cite{microdrivers, decaf, sud, lvds, lxds} choose to synchronize only a subset of structured object fields required for correct execution, primarily for improved performance. The required fields are normally computed via manual efforts~\cite{lxds, sud}, or static analyses~\cite{microdrivers, decaf, ksplit}. SFI based techniques~\cite{xfi, lxfi, bgi} allow developers to specify access capabilities at fine-grained levels, e.g., a byte or field levels, to prevent access on overshared data. Despite these efforts, addressing oversharing issues remains challenging. 

\subsection{Driver isolation security properties}
\label{subsec:drv_iso_security_properties}
\vspace{-0.5em}
Because they assume different threat models, prior driver isolation frameworks differ in the security properties that they enforce. Below, we collect a set of security properties that are enforced by those driver isolation frameworks. 

\trent{Still some issues.  In general, data integrity/secrecy may refer to illicit accesses (memory errors) or information flows.  We seem to consider P1 and P2 for illicit errors in some cases and information flows in others.  P4 allows illegal information flows with endorsement. P6 prohibits all illicit accesses.  P4 and P5 different forms of CFI, although the implied CFI policies should be made more explicit - which indirect calls are considered and what targets are allowed.}

\para{P1: Kernel data integrity} This property specifies that data stored in the memory of the kernel compartment cannot be directly updated by the driver compartments.

\para{P2: Kernel data confidentiality} This property specifies that data stored in the memory of the kernel compartment cannot be directly read by the driver compartments.

\para{P3: Interface data integrity} This property requires the integrity of data exchanged through the compartment interface. Even after P1 is enforced, the kernel and drivers still need to share data, as discussed in Section~\ref{subsec:iso_drv_shared_data}. If this interface data is corrupted to arbitrary values or to violate \emph{invariants} expected by the kernel, it can enable various attack vectors depending on how the kernel uses the data. As we will demonstrate in Section~\ref{sec:civ_overview}, failing to maintain this property can lead to multiple types of \vulName{}s related to the interface data.
\trent{This one is still too abstract.  I would say that it is an extended integrity check - like an endorser - which allows untrusted writes (unlike P1), but within a specification, which is effectively an endorser. Yongzhe: updated the writing to reflect this.} \gtan{Now we specify this as a property that requires the integrity of any data exchanged over the interface; I believe it's more clear now.}

\para{P4: Interface control flow integrity} A driver's control flow must remain confined within its code, except for invocations of predefined kernel interface functions and returns to the kernel call sites of driver functions. \trent{This one could come in multiple flavors since the set of legal indirect call targets could be dependent on the analysis to compute them.  However, I think you are implicitly making a distinction between limiting the indirect calls in the driver that may reach the kernel and the indirect calls that may not reach the kernel - as well as the target sets for the former.}

Properties P1-P3 focus on data flow, while P4 addresses control flow. Among these, P1 and P4 are enforced by all driver isolation frameworks, as they directly address the main goal of driver fault isolation across various threat models. They prevent drivers from corrupting kernel data directly and executing arbitrary kernel code. The remaining properties (P2 and P3) are primarily considered in frameworks that assume potentially malicious drivers. P2 prevents driver from reading sensitive information from kernel, while P3 ensures that all the shared data, even allowed to be legitimately updated by driver, cannot be corrupted. \trent{Still vague (Yongzhe): updated writing to make it more concrete.}, preventing the injection of malicious or erroneous values. 
In addition to aforementioned properties, some isolation frameworks extend security to the driver:

\para{P5: Driver control flow integrity}
This property is stronger than P4, specifying that the driver's control flow must adhere to a precomputed control flow graph. It constrains the usage of kernel interfaces, mitigating potential attacks such as use-after-free vulnerabilities \cite{assess_civ, bgi, dingo} that could arise from arbitrary invocation sequences of kernel-driver interface functions, even after P4 is enforced.

\para{P6: Driver memory safety}
Memory safety can be introduced to drivers in multiple ways: 1) utilizing memory-safe programming languages, such as Rust \cite{takeaway_native_udp_driver_in_rust, redleaf} or Java~\cite{decaf}, or 2) extending the C type system to prevent memory errors using compile-time or run-time checks~\cite{SafeDrive}. 

\section{Driver Isolation Frameworks} \label{sec:drv_iso_frameworks}
\begin{table*}[]
\centering
\resizebox{\textwidth}{!}{%
\begin{tabular}{|l|c|c|c|c|c|c|c|c|c|c|c|}
\hline
\multirow{2}{*}{\textbf{Framework}} &
  \multirow{2}{*}{\textbf{Year}} &
  \multirow{2}{*}{\textbf{Conference}} &
  \multirow{2}{*}{\textbf{Threat Model}} &
  \multirow{2}{*}{\textbf{Kernel Data Integrity}} &
  \multirow{2}{*}{\textbf{Kernel Data confidentiality}} &
  \multirow{2}{*}{\textbf{Interface Data Integrity}} &
  \multirow{2}{*}{\textbf{Interface CFI}} &
  \multirow{2}{*}{\textbf{Driver CFI}} &
  \multirow{2}{*}{\textbf{Mem Safety}} &
  \multirow{2}{*}{\textbf{Key Isolation technique}} &
  \multirow{2}{*}{\textbf{Platform}} \\ 
 &
   &
   &
   &
   &
   &
   &
   &
   &
   &
   &
   \\ \hline\hline
Nooks\cite{nooks} &
  2002 &
  SIGOPS &
  B &
  PT &
  N/A &
  N/A &
  WF &
  N/A &
  N/A &
  MMU &
  Linux \\ \hline
Driver reuse using via VM\cite{unmodified-device-drivers} &
  2004 &
  OSDI &
  B &
  GVA &
  GVA &
  N/A &
  Translation Module &
  N/A &
  N/A &
  VM (Xen) &
  Linux \\ \hline
SafeDrive\cite{SafeDrive} &
  2006 &
  OSDI &
  B &
  Mem Safety &
  Mem Safety &
  N/A &
  Mem Safety &
  N/A &
  Bounds Check &
  Language &
  Linux \\ \hline
XFI\cite{xfi} &
  2006 &
  OSDI &
  E &
  I &
  N/A &
  N/A &
  WF &
  CFI guards &
  N/A &
  SFI &
  Windows \\ \hline
Microdrivers\cite{microdrivers} &
  2007 &
  ASPLOS &
  B &
  Privilege Mode &
  Privilege Mode &
  SC &
  IPC &
  N/A &
  N/A &
  User-mode &
  Linux \\ \hline
TwinDrivers\cite{twindrivers} &
  2009 &
  ASPLOS &
  B &
  SVM &
  SVM &
  N/A &
  PE &
  N/A &
  N/A &
  VM (Xen) &
  Linux \\ \hline
Decaf\cite{decaf} &
  2009 &
  USENIX &
  B &
  Privilege Mode &
  Privilege Mode &
  SC &
  IPC &
  N/A &
  N/A &
  User-mode + Safe Lang &
  Linux \\ \hline
Protecting OS from vul drvs\cite{protecting_comm_kernel_from_vulnerable_driver} &
  2009 &
  ACSAC &
  E &
  Privilege Mode &
  Privilege Mode &
  SC &
  IPC &
  N/A &
  N/A &
  User-mode + Daikon &
  Linux \\ \hline
BGI\cite{bgi} &
  2009 &
  SOSP &
  E &
  I &
  Cap &
  Cap &
  WF &
  CFI guards &
  byte-granularity CAP &
  SFI &
  Windows \\ \hline
SUD\cite{sud} &
  2009 &
  USENIX &
  E &
  Privilege Mode &
  Privilege Mode &
  N/A &
  IPC &
  N/A &
  N/A &
  User-mode &
  Linux \\ \hline
HUKO\cite{huko} &
  2011 &
  NDSS &
  E &
  MAC &
  MAC &
  N/A &
  PE &
  N/A &
  N/A &
  VM (Xen) &
  Linux/Windows \\ \hline
LXFI\cite{lxfi} &
  2011 &
  SOSP &
  E &
  CAP &
  CAP &
  N/A &
  CAP &
  CAP + Shadow Stack &
  N/A &
  SFI &
  Linux \\ \hline
SIDE\cite{side} &
  2013 &
  DSN &
  B &
  Privilege Mode &
  Privilege Mode &
  N/A &
  WF &
  N/A &
  N/A &
  User-mode &
  Linux \\ \hline
LXDs\cite{lxds} &
  2019 &
  USENIX &
  B &
  GVA &
  GVA &
  SC &
  IPC &
  N/A &
  N/A &
  VM (KVM) &
  Linux \\ \hline
LVDs\cite{lvds} &
  2020 &
  VEE &
  B &
  GVA &
  GVA &
  SC &
  IPC &
  N/A &
  N/A &
  VM (Bareflank) &
  Linux \\ \hline
KSplit\cite{ksplit} &
  2022 &
  OSDI &
  B &
  GVA &
  GVA &
  SC &
  RPC &
  N/A &
  N/A &
  VM (Bareflank) &
  Linux \\ \hline
DriverJar\cite{driverjar} &
  2023 &
  DAC &
  E &
  Trampoline &
  N/A &
  N/A &
  PE &
  N/A &
  N/A &
  Hardware watchpoint &
  Arm-based \\ \hline
Sfitag\cite{sfitag} &
  2023 &
  Asia CCS &
  E &
  Tagging (MTE) &
  Tagging (MTE) &
  N/A &
  WF + tagging &
  N/A &
  N/A &
  Memory tagging &
  Arm-based \\ \hline
\end{tabular}%
}
\caption{Existing driver isolation frameworks, their threat models (B: Buggy, E: Exploitable), and enforcement techniques for security properties P1-P6. Techniques are abbreviated as SC (Selective copy), PE (Predefined entries), I (Instrumentation), MAC (Mandatory access control), SVM (Software virtual memory), GVA (Guest virtual address), CAP (Capability), and WF (Wrapper function). N/A indicates the property is not enforced.
}
\label{tab:drv_iso_frameworks_summary}
\end{table*}

In this section, we examine various driver isolation techniques and analyze the security properties they enforce. We delve into the implementation details of these isolation techniques to understand how they enable specific security properties. Table~\ref{tab:drv_iso_frameworks_summary} provides a comprehensive overview of the frameworks that we have investigated. For each security property, we indicate whether it is enforced, and, if so, we describe the corresponding implementation mechanism.

\subsection{Software-based Fault Isolation (SFI)}
\label{subsec:drv_iso_classification_sfi}
\vspace{-0.5em}
SFI-based techniques offer driver isolation without hardware support by establishing distinct protection domains within a shared address space using static analysis and inline software \textit{guards} for runtime checks \cite{Wahbe93SFI, Tan17SFI}. Notable examples include XFI \cite{Erlingsson06XFI}, BGI \cite{Castro09BGI} for Windows, and LXFI \cite{Mao11LXFI} for Linux. Recent advancements focus on enhancing SFI performance using hardware features like ARM MTE \cite{sfitag} and PAC~\cite{hakc}.

To enforce P1, SFI instruments driver memory instructions to restrict access within the driver's domain or explicitly granted memory ranges, configured by a trusted monitor or capabilities~\cite{Mao11LXFI}. P4 is enforced by instrumenting \textit{jump} instructions, restricting targets to the driver's domain or preconfigured addresses. Violations trigger a fault and module restart. XFI and LXFI enforce additional properties, assuming potentially malicious drivers. XFI incorporates a verifier for P5 but don't check interface data, thus fail to enforce P3~\cite{Erlingsson06XFI}. LXFI addresses this with \textit{API integrity} via developer-supplied annotations for P3 and ensures these checks cannot be bypassed by enforcing P5~\cite{Mao11LXFI}. LXFI achieves this by using a combination of capability and shadow stack. BGI assumes a buggy driver model, implementing a superset of XFI's runtime checks while assuming unaltered control flow, balancing security and performance for bug-focused scenarios. Since BGI allows assigning capability at byte-granularity level, it can theoratically achieves memory safety if all the memory access rights are configured correctly by the developer. In general, SFI-based architectures ignore instrumentation on memory reads for performance, thus not enforcing P2.

\subsection{Language-based driver isolation}
\vspace{-0.5em}
Language-based driver isolation enforces memory safety (P6) within drivers using type systems or safe programming languages~\cite{SafeDrive, takeaway_native_udp_driver_in_rust, redleaf}. SafeDrive~\cite{SafeDrive} uses a 
type system called \textit{Deputy} to prevent pointer bounds errors through compile-time and runtime checks, relying on annotations that can be partially inferred and supplemented by developers to ensure the safety of memory access types, including pointer bounds and union selectors.
With complete annotations, SafeDrive detects and prevents all memory errors within the driver, enforcing P6. Consequently, it achieves key security properties of other driver isolation frameworks: P1 and P2 are enforced because the driver cannot directly corrupt or read kernel private data through memory bugs, and P4 is easily enforced since the driver cannot corrupt function pointers or return addresses, ensuring communication with the kernel only through predefined interface functions. However, SafeDrive does not enforce the remaining security properties. Similarly, rewriting drivers in safe languages like Rust~\cite{takeaway_native_udp_driver_in_rust, redleaf} achieves comparable isolation guarantees.

\subsection{User-mode drivers}
\label{subsec:user-mode-drv}
\vspace{-0.5em}
User-mode drivers run driver code in userspace processes, leveraging hardware privilege separation (Ring 0/3) to prevent direct driver access to kernel memory, inherently enforcing P1~\cite{microdrivers, decaf, sud}. Communication between userspace drivers and the kernel is performed through message-based channels, such as system calls in Microdrivers~\cite{microdrivers} and Decaf~\cite{decaf}, or RPC in SUD~\cite{sud}. This restricted interface approach inherently enforces P4, limiting drivers to invoke kernel code through predefined interfaces.

In user-mode drivers, data is passed via messages, maintaining separate object copies in the driver and kernel, synchronized during cross-domain invocations. However, deep copying large and complex kernel objects at each domain crossing incurs high runtime overhead. Therefore, object synchronization across boundaries is typically selective, focusing on fields critical for correct execution. Microdrivers~\cite{microdrivers} and KSplit~\cite{ksplit} employ static analysis to automate the identification of necessary fields.

Although primarily motivated by performance, object copying also facilitates the enforcement of P1 and P2 by not synchronizing kernel private data. P3 is not enforced for most of user-mode driver isolation projects, but an extension to the Microdriver architecture by Butt et al.~\cite{protecting_comm_kernel_from_vulnerable_driver} infers and checks data invariants during domain crossings using the dynamic invariants inference tool Daikon~\cite{daikon}.
User-mode driver architectures generally do not explicitly address P5 or P6. 

\subsection{Page table switching}
\label{subsec:pt-iso}
\vspace{-0.5em}
Multiple projects utilize MMU to establish distinct page tables for different compartments. Nooks~\cite{nooks} employs page tables to enforce isolation, allowing kernel code unrestricted access to the driver's memory while restricting the driver's write access to kernel private data (P1). However, to synchronize kernel updates, Nooks allows the driver to read the kernel's page table and copy the data to its compartment, therefore violating P2 (kernel data confidentiality). Nooks provides wrapper functions for cross-domain invocations but do not enforce control flow integrity. Invocations from the driver to kernel are not guaranteed to jump to these wrapper functions, and thus both P4 and P5 are not enforced, allowing attacks like Page-Oriented Programming~\cite{pop} to hijack control flow. Nooks does not enforce P3 as it doesn't limit or check data passed through the interface.

SIDE~\cite{side} uses MMU to run drivers within the kernel address space, allocating a specific region for each driver with user-level privilege (ring 3). The isolated driver has its private stack and heap within this region. SIDE enforces P1 and P2, as unprivileged driver access to the kernel triggers a ring exception, invoking a handler to validate the access. Control transfers across isolation domains generate ring exceptions, allowing the handler to verify the transfer's legitimacy (P4). However, SIDE does not address P3, P5, or P6.

\subsection{Virtualization}
\label{subsec:vm-based-iso}
\vspace{-1em}
Virtualization provides a natural way to run untrusted code, and various driver isolation frameworks based on virtualization have been developed~\cite{unmodified-device-drivers, lxds, lvds, twindrivers, huko}. In an early demonstration, LeVasseur et al.~\cite{unmodified-device-drivers} used Xen virtualization to run buggy device drivers along with a native OS in an unprivileged domU VM, with a translation module added to the driver's OS to serve as a server for external requests. This design achieves P1 and P2 due to VM isolation, and P4 by using a specified set of functions for communication between the driver and the external system.

Twindrivers\cite{twindrivers}, also built on Xen, employs an isolation boundary similar to \cite{microdrivers}, running performance-critical functions within the hypervisor domain and the remaining functions in the dom0 domain. It uses \textit{software virtual memory (SVM)} to restrict hypervisor instance accesses to a single object in the dom0 instance's address space, enforcing P1, P2, and P4 through predefined upcalls and hypercalls.

Other hypervisors, such as KVM in LXDs~\cite{lxds} and Bareflank in LVDs~\cite{lvds}, have also been used for driver isolation. LVDs optimize cross-VM communication using Intel VMFUNC instructions, avoiding the overhead of trapping to the hypervisor on every interaction. In these approaches, the memory accessible by the driver is limited to the VM virtual address space, and control transfers out of the driver must go through the hypervisor, enforcing P1, P2, and P4. However, it's worth noting that some of these projects perform selective copying, such as LVDs and KSplit. This approach may reduce the fields that required integrity checking (P3). However, P3 is generally not considered enforced. 

\subsection{Memory tagging}
\label{subsec:mem-tagging-iso}
\vspace{-0.5em}

Memory tagging augments code and data with \textit{tags}, which function as security metadata on memory. These tags are used to enforce memory access policies~\cite{tag_guide}. At a high level, memory tagging establishes isolation domains similar to SFI-based techniques. However, memory access operations are checked via hardware with much lower overhead (with an average of less than 5\% compared to native systems). 

In the context of driver isolation, the only work we have found that utilizes memory tagging for driver isolation purposes is Sfitag~\cite{sfitag}, which utilizes ARM MTE. Sfitag generally enforces the same security properties as SFI-based techniques, such as P1 and P4. Additionally, because of the lower overhead of checking memory access operands using hardware support, Sfitag is also able to instrument memory read instructions and thus enforces P2, compared to SFI-based approaches. However, ARM MTE has been demonstrated to suffer from speculative attacks~\cite{tiktag}, hindering the security guarantees of Sfitag.

\para{Summary} To summarize existing works, almost all driver isolation projects enforce P1 and P4. These two properties are essential for ensuring that a buggy or exploitable driver cannot interfere with the kernel's correct execution or execute arbitrary kernel functions. Projects that assume an exploitable driver also commonly consider P2, as the driver could potentially read sensitive kernel information. However, P3 is rarely enforced, with only a few projects~\cite{protecting_comm_kernel_from_vulnerable_driver, lxfi} partially enforcing this property through checks at domain crossings, mainly due to the difficulty of inferring data invariants for integrity checks. P5 is rarely considered, except for SFI techniques. The main reason such projects need P5 is to ensure that the added instrumentations cannot be bypassed by the driver. P6 can be enforced by language-based approaches, e.g., by performing driver rewriting using safe programming language. 

\section{Compartment Interface Vulnerabilities}
\label{sec:civ_overview}
\vspace{-0.5em}
Driver isolation effectively confines the memory accesses of potentially buggy drivers. However, even after isolation, a compromised or malicious driver can attack the kernel by misusing the kernel/driver interface. These potential attacks, termed \textit{Compartment Interface Vulnerabilities} (CIVs) by previous research~\cite{assess_civ, CIVScope} are specific to compartmentalized applications when assuming a malicious compartment.

Researchers have long been aware of the potential for attacks targeting the kernel/driver interface in compartmentalized systems. Early microkernel designs like MINIX 3 demonstrated that a buggy isolated module can violate its IPC protocol and cause issues such as deadlocks~\cite{counter_ipc_threats_multiserver}. LXFI~\cite{lxfi} addressed interface function argument integrity using capabilities, and Butt et al.~\cite{protecting_comm_kernel_from_vulnerable_driver} attempted to infer and enforce data invariants to prevent drivers from supplying corrupted data to the kernel. However, most existing driver isolation frameworks do not address \vulName{}s, as illustrated in Section~\ref{subsec:drv_iso_security_properties}, where security properties that address the misuse of interfaces and shared data, such as P3 and P5, are not widely enforced. Based on an examination of existing literature, particularly RLBox \cite{RLBox} and Conffuzz \cite{assess_civ}, this section presents a taxonomy of \vulName{}s, broadly categorizing them into three types: (1) \textit{shared data}, (2) \textit{concurrency}, and (3) \textit{control transfer}. \trent{Concurrency impact should be discussed earlier. One of the properties - address by microdrivers and KSplit and ...}

\subsection{CIV Taxonomy} \label{subsec:drv_iso_civ_taxonomy}
The taxonomy we consider for this study is presented in Table~\ref{tab:civ_taxonomy}. It is based on previous work \cite{RLBox, assess_civ, CIVScope}, with an extension of five new \vulName{}s, highlighted in \textbf{bold} text. We also provide citations for the \vulName{}s that have been studied by previous work. To avoid redundant discussion, we provide code examples only for the new \vulName{}s. 
In addition, we note that while we have made our best effort to summarize and extend the existing taxonomy, we do not claim that this list covers all possible \vulName{}s. As research in this area progresses, future work may discover and add newer \vulName{}s to the taxonomy. Next, we discuss each category and the new \vulName{}s in detail.

\begin{table*}[h]
\centering
\begin{tabular}{|p{0.15\textwidth}|p{0.2\textwidth}|p{0.15\textwidth}|p{0.35\textwidth}|}
\hline
\textbf{High-level Category} & \textbf{CIV Category} & \textbf{Subcategory} & \textbf{Detailed Instances} \\ \hline
\multirow{12}{*}{\textsc{Shared Data}} & \multirow{2}{*}{Shared data leakage} & \multirow{2}{*}{} & Leaking non-pointer values \cite{assess_civ, CIVScope} \\ \cline{4-4} 
 & & & Leaking pointer values \cite{assess_civ} \\ \cline{2-4} 
 & \multirow{10}{*}{Shared data corruption} & \multirow{6}{*}{Memory safety} & Corrupted pointer value \cite{assess_civ, CIVScope} \\ \cline{4-4} 
 & & & Corrupted pointer offset or buffer index \cite{assess_civ} \\ \cline{4-4} 
 & & & \textbf{Corrupted union type selectors} \\ \cline{4-4} 
 & & & Parameters used in kernel memory API \cite{CIVScope} \\ \cline{4-4} 
 & & & Corrupted string \cite{assess_civ} \\ \cline{4-4} \cline{3-4} 
 & & \multirow{3}{*}{Decision making} & Corrupted guard (data attack) \cite{assess_civ} \\ \cline{4-4} 
 & & &  Return wrong/invalid error code \cite{dingo} \\ \cline{4-4}
 & & &  \textbf{Corrupted loop condition} \\ \cline{3-4}
 & & \multirow{2}{*}{Arithmetic error} & Divided by zero \cite{printfuzz} \\ \cline{4-4} 
 & & & Integer overflow/underflow \cite{printfuzz} \\ \hline\hline
\multirow{3}{*}{\textsc{Concurrency}} & \multirow{3}{*}{Race condition} & \multirow{3}{*}{} & Corrupted synchronization primitive \cite{assess_civ} \\ \cline{4-4} 
 & & & Callback state exchange \cite{RLBox} \\ \cline{4-4} 
 & & & Shared memory TOCTTOU \cite{assess_civ, RLBox} \\ \hline\hline
\multirow{5}{*}{\textsc{Control Transfer}} & \multirow{2}{*}{Interface bypass} & \multirow{2}{*}{} & Corrupted function pointer invoked by driver \cite{assess_civ} \\ \cline{4-4} 
 & & & Corrupted function pointer invoked by kernel \cite{assess_civ} \\ \cline{2-4} 
 & \multirow{3}{*}{Interface temporal violation} & \multirow{3}{*}{} & \textbf{Sleep in an atomic context} \\ \cline{4-4} 
 & & & \textbf{Lock and never unlock} \\ \cline{4-4} 
 & & & \textbf{Unbalanced allocation/deallocation} \\ \hline
\end{tabular}
\caption{CIV Taxonomy. The first column presents the high-level categories; the second column presents the main CIV categories; the third column further classifies the CIVs into more specific subcategories; the fourth column lists specific types of vulnerabilities that fall under a subcategory, along with references to relevant literature.}
\label{tab:civ_taxonomy}
\end{table*}

\subsection{Shared data CIVs}
\label{subsec:shared_data_civ}
As discussed in Section~\ref{subsec:iso_drv_shared_data}, shared data refers to the data that can be accessed by both the driver and the kernel after isolation. These include arguments, return values of interface functions, and global variables. 

\subsubsection{Shared data leakage}
\label{subsec:shared_data_leakage_civ}
Shared data leakage \vulName{}s can expose kernel
confidential information to an isolated driver. Commonly, such leakage can take two forms:
\begin{itemize}
    \item \textbf{Data oversharing}: Data leakage can occur when there is a lack of fine-grained access control on large and complex aggregate type objects. For example, it is common that an entire struct is shared while only a subset of its fields is needed by the receiving domain.
    
    \item \textbf{Uninitialized data}: Uninitialized shared objects allocated by the kernel may contain kernel sensitive data that can leak to the driver compartment during kernel-driver synchronization. This can occurs through either incomplete initialization after object allocation or compiler-added padding bytes for alignment purposes~\cite{assess_civ, unisan}.
\end{itemize}

Leaked data can be of two types: (1) a pointer type, where leaked pointer values can expose address layout information, allowing the subversion of memory layout randomization techniques; and (2) a non-pointer type, which may contain sensitive kernel information, such as cryptographic keys, authentication tokens, or user data.

\subsubsection{Shared data corruption}
\label{subsec:shared_data_corruption_civ}
Shared data corrupted by malicious drivers can affect kernel operations when used in critical tasks such as memory API calls~\cite{CIVScope}. Conffuzz's \vulName{} taxonomy classifies shared data corruption based on data type, e.g., corrupted pointers, indices, and objects like strings. However, this classification does not fully convey the end kernel uses of the corrupted data. We classify shared data corruption \vulName{}s based on the kernel operations that use corrupted data, categorizing them as: (1) \textit{memory safety} violations, (2) \textit{decision making} violations, and (3) \textit{arithmetic errors}.

\para{Memory safety} Memory safety \vulName{}s occur when the kernel uses driver-corrupted data in memory operations. For example, a corrupted shared pointer or offset can cause corrupted kernel memory~\cite{CIVScope}, enabling advanced attacks like DUI~\cite{dui}, which achieve the equivalent of controlling kernel read/write primitives. 

\para{Decision making} Attackers can corrupt shared data used in kernel execution control flow decisions, manipulating the control flow and steering it to paths containing sensitive kernel operations.

\para{Arithmetic error} Arithmetic operations on corrupted data can lead to kernel faults. For example, a malicious driver supplying a zero divisor can trigger a divide-by-zero exception and subsequent kernel panic. Integer overflow or underflow can generate unexpected values, potentially bypassing security checks or causing API misuse when propagated.

\para{New shared data corruption instances}
Next we discuss two new instances of shared data corruption CIVs: (1) corrupted type selectors in unions, and (2) corrupted loop condition.  

\begin{listing}[t]
\begin{minted}{c}
union acpi_object {  @\phantomsection\label{lst:union:start}@
  acpi_object_type type;
  struct {acpi_object_type type; u64 value;} integer;
  struct {acpi_object_type type; ...; u8* pointer} buffer;
  struct {..., union acpi_object *elements;} package;
  ...
}; @\phantomsection\label{lst:union:end}@

acpi_status acpi_extract_package(union acpi_object *package, ...) {
  ...
  for (i = 0; ... ; i++) {
    union acpi_object *element=&(package->package.elements[i]);   
    switch (element->type) {
      case ACPI_TYPE_BUFFER:
      ...
    }
  }
}
\end{minted}
\caption{Example of corrupted union type selector.}
\label{lst:type-confusion}
\end{listing}

\emph{Corrupted union type selector} The kernel often relies on tagged unions and void pointer types to implement polymorphism, interpreting the data-structure type based on context or an associated tag value. In a type confusion attack, the attacker confuses the kernel code into accessing a data structure using the wrong interpretation. For example, the \lstinline{acpi_power_meter} driver uses a tagged union \lstinline{acpi_object} to represent various ACPI objects (\autoref{lst:type-confusion}). The driver invokes \lstinline{acpi_extract_package} to extract objects of a particular type. However, a malicious driver could corrupt the extracted element's type field, causing the kernel to misinterpret the memory layout (line 12). It could trick the kernel into treating an \lstinline{integer} struct as a \lstinline{buffer} struct. The \lstinline{buffer} struct's third field, \lstinline{pointer}, is an address. By crafting the \lstinline{integer} struct's \lstinline{value} field with a specific address and corrupting the \lstinline{type} field, the malicious driver can trick the kernel into performing read/write operations at arbitrary memory locations via the \lstinline{value} field.

\emph{Corrupted loop condition} These can lead to various security vulnerabilities. A basic example is a DoS attack, where affecting the loop condition to cause a large number of loops can cause the kernel to have degraded performance or even become unresponsive. A more critical vulnerability occurs when an attacker can manipulate a loop termination condition that controls both the iteration count and buffer indexing. This scenario can result in a buffer overflow attack. For instance, in the \lstinline{mgag2000} driver, an attacker can corrupt the \lstinline{n_layers} field, which controls both the loop iterations and array access to \lstinline{mci->layers}, potentially leading to a buffer overflow vulnerability, as shown in ~\autoref{lst:corrupted_loop_condition}.

%
%
%

\begin{listing}[t]
\begin{minted}{c}
void edac_mc_handle_error(..., struct mem_ctl_info *mci, ...) {
    for (i = 0; i < mci->n_layers; i++) {
	   if (pos[i] >= (int)mci->layers[i].size)
    }
    ...
}
\end{minted}
\caption{Example of buffer overflow with corrupted loop condition.}
\label{lst:corrupted_loop_condition}
\end{listing}

\subsection{Race conditions} \label{subsec:race_condition_civ}
Race conditions involve concurrent accesses to shared data. We consider three types of concurrency \vulName{}s.

\para{TOCTTOU}
Time-of-Check-to-Time-of-Use vulnerabilities (TOCTTOU) can occur when an isolated driver modifies shared data between a kernel's check and use. This creates a window for the driver to manipulate the checked value post-validation. To address the TOCTTOU issue, the check and the use of the check must be atomic. For example, RLBox~\cite{RLBox} provides a mechanism for making a copy when a compartment receives data and ensures that all checks and uses are performed on the copied data.

\para{Shared Lock Corruption}
An isolated driver with the ability to modify shared lock values can compromise atomic regions dependent on these locks. This represents a specialized form of shared data corruption~\cite{assess_civ}. 

\para{Callback State Exchange}
RLBox~\cite{RLBox} identifies a multithreading attack where a malicious compartment provides corrupted object instances to multiple threads in the trusted compartment. For example, a compromised driver could supply identical device objects to different threads handling distinct devices, leading to unexpected race conditions in kernel API invocations.

\subsection{Control transfer CIVs}
\label{subsec:control_transfer_civ}
\vspace{-0.5em}
Control flow transfer CIVs fall into two categories: (1) {\em interface bypass} and (2) {\em interface temporal violations}.

\subsubsection{Interface bypass}
Driver isolation frameworks restrict the driver and kernel to interact through a set of predefined interfaces. In the kernel-driver context, interface functions are accessed via function pointers. If an isolated driver can corrupt these function pointers, it may either use a corrupted function pointer to invoke arbitrary kernel functions or trick the kernel into jumping to arbitrary code locations when invoking driver callbacks through these pointers. However, if P4 (interface control flow integrity) is enforced, these attacks would be prevented even with corrupted function pointers. We discuss this assumption further in Section~\ref{subsec:quantify_control_transfer_civ} when addressing control transfer \vulName{} quantification.

\subsubsection{Interface temporal violation}
An interface temporal violation is normally caused by calling interface functions in a wrong order. That is, although a driver may invoke only allowed kernel interface functions, doing so in the wrong order is possible without further defenses.  While there are many kinds of violations, we discuss three such \vulName{} types in this study: (1) Sleep in atomic context, (2) Lock and never unlock, and (3) Unbalanced allocation/deallocation.

\para{Sleep in atomic contexts (SAC)} 
This is a \vulName{} type specific to the kernel context~\cite{dsac}. Atomic contexts, such as spinlock-protected regions or interrupt handlers, require the driver to complete its operations promptly and without blocking. However, if a driver invokes a sleepable kernel interface while holding a spinlock, such as calling kernel memory allocation functions (e.g., \lstinline{kmalloc()}) without passing the \lstinline{GFP_ATOMIC} flag, the system can deadlock or crash.
Such vulnerabilities can be detected using static analysis~\cite{dsac}.

\para{Lock and never unlock} A malicious driver can  hold shared locks required by other kernel threads for making progress, causing system-wide hangs or denial-of-service attacks.

\para{Unbalanced allocation/deallocation}
Drivers are responsible for correctly manage object lifetime via kernal memory management APIs, e.g., kmalloc/kfree. A malicious driver can intentionally create memory leaks by bypassing calls to \texttt{free} on allocated objects. This can potentially lead to DoS attacks by depleting system memory. We show an instance of this \vulName{} in Appendix~\ref{app:failed_shared_data_corruption} (  Listing~\ref{lst:vulnerable-bonding-driver}).

%% file: tex/threat_model.tex
%% file: tex/meth.tex
\section{Evaluating Driver Isolation Effectiveness}
\label{sec:methodology}
\vspace{-0.3em}
Our discussion thus far has covered existing driver isolation techniques and the classes of \vulName{}s that can compromise driver isolation security. However,  several key questions remain unanswered:
\begin{enumerate}
    \item Which \vulName{}s remain possible under driver isolation?
    \item How prevalent are these \vulName{}s?
    \item How certain security properties  can be leveraged to mitigate \vulName{}s? 
\end{enumerate}

To answer these questions, we perform our evaluation under different threat models. At a high level, we first begin with a baseline isolation model common to most driver isolation frameworks to answer Q1 and Q2. Then, we explore how extra properties such as P3, P5 and P6, can be applied to reduce the number of \vulName{}s.

\subsection{Threat models}\label{subsec:eval_threat_model}
\vspace{-0.6em}
We define a baseline threat model for an isolated driver framework that enforces the following security properties: (1) P1: kernel data integrity, (2) P2: kernel data confidentiality, and (3) P4: interface control flow integrity. In this model, we also assume selective sharing of data, where only necessary fields in objects are synchronized. This partially enforces P3 by avoiding synchronizing unnecessary data; however, we do not assume a defense that infers and validates invariants of interface data, which is an unsolved problem.
\trent{This prior sentence should describe a property.  Need to define this - may need multiple levels of P3, since we do not achieve that completely.} \gtan{I tried to clarify a bit above.}
This driver isolation model captures the majority of driver isolation frameworks that aim to confine a buggy driver by controlling data sharing. 
We consider that an exploitable driver contains bugs that allow an attacker to execute arbitrary code within the driver. Consequently, all driver private data and shared data can be read and modified by the driver, the control flow within the driver can be hijacked, and arbitrary code can be executed, enabling the attacker to jump to any driver location and invoke any kernel interface function.

To investigate how \vulName{}s may be mitigated by enforcing additional security properties, we compare numbers of \vulName{}s in the baseline model with the numbers when a security property is enforced on top of the baseline. This approach allows us to understand the effect of enforcing a security property on \vulName{}s. 


\subsection{Methodology of quantifying \vulName{}s} \label{sec:method_quantifying_attack_surface}
In this section, we present our methodology for quantifying \vulName{}s at the kernel-driver isolation boundary.
We summarize the \vulName{} classes, and the corresponding metrics in 
\autoref{tab:civ_quantification_metrics}.  
Our quantification methodology relies on static analysis, which identifies CIVs that can potentially be exploited given our threat model. Since static analysis may produce false positives, we will present how we further validate our results using manual analysis and proof of vulnerabilities in Section~\ref{subsec:eval_civ_classification}. 
We next explain the detailed method for quantifying each \vulName{} class. 

\begin{table}[h]
\centering
\small
\resizebox{\columnwidth}{!}{%
\begin{tabular}{|l|l|p{7cm}|}
\hline
\textbf{Category} & \textbf{CIV Class} & \textbf{Metrics} \\
\hline\hline
\multirow{2}{*}{Shared data} & Data leakage & Count of total fields - Count of shared fields \\
\cline{2-3}
 & Data corruption & Count of taint paths \\
\hline
Concurrency & Race & Count of corruptible shared locks \\
\hline
\multirow{3}{*}{Ctrl transfer} & Sleep in atomic context & Count of (spinlock, sleepable func) pairs \\
\cline{2-3}
 & Lock never unlock & Count of lock/unlock pairs on shared locks \\
\cline{2-3}
 & Unbalanced allocation & Count of allocation/deallocation pairs \\
\hline
\end{tabular}%
}
\caption{CIV classes and their metrics.}
\label{tab:civ_quantification_metrics}
\end{table}

\vspace{-0.5em}
\subsection{Quantify shared data leakage CIVs}\label{subsec:shared_data_confidentiality_quantify_method}
As described in Section~\ref{subsec:shared_data_leakage_civ}, kernel sensitive information can be leaked due to overshared or uninitialized data. For our evaluation, we focus on measuring the degree of overshared data and omit the quantification of uninitialized data for two reasons: (1) it is hard to determine uninitialized data due to compiler-padding and (2) in certain object-copying based techniques, access is granted to only the private object copy in the driver and thus the driver cannot access uninitialized data in the kernel's copy to learn confidential data.

To measure the degree of oversharing, we utilize the \textit{shared field analysis} proposed in KSplit~\cite{ksplit} to identify those struct fields whose states are required for correct execution. KSplit's analysis is built on top of the algorithm used by Microdrivers~\cite{microdrivers}, and further improves precision by determining shared fields. For example, in Microdriver, an interface parameter field updated by the driver will be synchronized to the kernel. However, if the field is used by only the driver but not the kernel, it is considered private to the driver and KSplit can determine that it is unnecessary to synchronize the field.

Our algorithm takes all structure-typed interface function parameters, return values, and shared global variables as input, and computes the following metrics: (1) the total number of structure fields if deep copy is used, (2) the number of accessed structure fields, and (3) the number of shared and accessed structure fields (via shared field analysis). The difference between the field numbers computed by (1) and (3) captures the degree of oversharing.

\subsection{Quantify shared data corruption \vulName{}s}
\label{subsec:shared_data_integrity_quantify_method}
\vspace{-0.5em}
This quantification procedure consists of two steps. The first step is a taint analysis because most of the risky vectors involve corrupting interface data (taint source) to affect a kernel operation (taint sink). Its output is a set of taint paths from the sources to the sinks. This methodology is similar to CIVScope~\cite{CIVScope}, but extended to a more complete CIV classification. 
Similar to CIVScope, our taint analysis faces the taint path explosion challenge, as not all of the sinks can be assumed to be exploitable given the possibility of condition checks on the taint path. Therefore, in the second step, we prune taint traces to prioritize those with path conditions that are \textit{controllable} by the attacker. This step allows us to improve the precision of the attack surface quantification.
\begin{table*}[h]
\centering
\scriptsize
\begin{tabular}{p{0.35\textwidth}p{0.15\textwidth}p{0.15\textwidth}p{0.25\textwidth}}
\toprule
\multirow{2}{*}{\textbf{\textsc{Shared Data Integrity \vulName{} Classes}}} & \multicolumn{3}{c}{\textbf{\textsc{Analysis Description}}} \\ 
\cmidrule(l){2-4}
& \textbf{Src} & \textbf{Propagation} & \textbf{Sink} \\
\midrule
\textbf{MEM1}: Corrupted pointer value & Pointer & Data & Ptr dereference operations \\ 
\textbf{MEM2}: Corrupted pointer offset/buffer index & Scalar & Data & Buffer access or ptr arith operations \\  
\textbf{MEM3}: Corrupted union type selector & Tagged union & Data & Uses of the union data \\ 
\textbf{MEM4}: Parameters in kernel memory APIs & All & Data & Sensitive kernel APIs \\
\textbf{MEM5}: Corrupted string & String & Data & String operations \\  
\midrule
\textbf{DM1}: Corrupted guard & All & Data & Branch instructions \\
\textbf{DM2}: Return invalid/wrong error code & Driver return val & Data & Branch instructions \\
\textbf{DM3}: Corrupted loop condition & Scalar & Data/Control & Buffer access inside loop body and loop condition\\  
\midrule
\textbf{AE1}: Divided by zero & Scalar & Data & Divide operations \\ 
\textbf{AE2}: Integer overflow/underflow & Scalar & Data & Overflowing arithmetic ops \\ 
\bottomrule
\end{tabular}
\caption{\vulName{}s classification. \textbf{Src} column shows the data type passed to interface functions that are used as taint source (e.g. "Pointer" and "All" refer to pointer/all data passed to interface functions, respectively). \textbf{Propagation} column shows control/data dependency. \textbf{Sink} column shows the sink of the taint analysis.
}
\label{tab:data_corruption_taint_details}
\end{table*}

\para{Taint analysis} \label{subsubsec:taint} Algorithm~\ref{alg:taint_analysis} presents the taint analysis, and Table~\ref{tab:data_corruption_taint_details} lists the included source/sink types.
The analysis takes all the arguments and return values of interface functions as well as global variables as taint sources, and a set of specified kernel operations as taint sinks. It outputs a set of taint paths from the sources to the sinks. 
Our taint analysis propagates taints on a field-sensitive, path-, and context-insensitive, interprocedural \emph{program dependence graph} (PDG) in LLVM~\cite{LiuTJ17Ptrsplit}. The PDG implementation captures the data and control dependencies among LLVM IR instructions and variables. It utilizes the SVF alias analysis~\cite{sui2016svf} to capture intraprocedural pointer aliasing. Interprocedural aliasing, on the other hand, is captured by interprocedural dependence edges in the PDG. Although the PDG lacks path and context sensitivity, it allows us to identify all potential risky operations that could operate on tainted data.

\newcommand{\taintTr}{\mathit{Tr}}
\begin{algorithm}[h]
\small
\SetAlgoLined
\KwIn{Taint sources $S$; PDG $G$}
\KwOut{Taint traces from sources to sinks}
\SetKwFunction{Ftaint}{TaintAnalysis}
\SetKwProg{Fn}{Function}{:}{}
\Fn{\Ftaint}{
    $T$ = risky kernel operations (Table~\ref{tab:data_corruption_taint_details})\;
    $\taintTr = \{\}$\;
    \ForEach{$s \in S$}{
        Propagate taint from $s$ along data-dependency edges of $G$\;
        \If{node $n$ becomes tainted and $n \in T$}{
            $\taintTr \gets \taintTr \cup \{$trace from $s$ to $n\}$\;
        }
    }
    \KwRet $\taintTr$
}
\vspace{0.5em}
\caption{Taint Analysis}
\label{alg:taint_analysis}
\end{algorithm}

\para{Finding likely attackable paths with pruning} 
\label{subsec:pruning_taint_trace}
A taint trace produced by our analysis indicates a potentially risky operation
that can be affected by values controlled by the attacker. However,
the risky operation may be protected by various checks; consequently,
whether the risky operation is exploitable depends on whether the
attacker can bypass those checks. Thus, to identify paths that are likely
to be exploited, we employ the following two heuristics: 

\begin{enumerate}

\item there is no check in the taint trace,
\item or there are checks in the taint trace, but none of them  directly checks the data used in the sink. 
\end{enumerate}

The first heuristic identifies those taint traces that are easily exploitable by the attacker, because the attacker can reach the sink without going through any check. The second heuristic identifies those traces whose sinks are guarded by some checks but those checks may not be effective as they do not directly check data used by the sinks. If a taint trace meets one of these heuristics, we preserve the trace. However, these heuristics may produce both false positives and false negatives. We provide evaluation of these heuristics in the evaluation section (Section~\ref{sec:experiment}).

\subsection{Quantify concurrency \vulName{}s} \label{subsec:quantify_race_condition}
We focus on quantifying corrupted shared locks, a specific type of concurrency-related \vulName{}. To identify shared data susceptible to this type of CIV, we extend KSplit's shared data analysis~\cite{ksplit}. This is a field-sensitive analysis that computes shared data based on the accesses in both domains. And then, we identify data that are used within lock APIs, to obtain the shared lock instances.  

\subsection{Quantify control transfer \vulName{}s}
\label{subsec:quantify_control_transfer_civ}
As discussed, control flow transfer \vulName{}s are categorized into two categories. The first one involves bypassing predefined interfaces by corrupting function pointers. In our baseline model, we assume that P4 (interface control flow integrity) is enforced. In an exploitable driver threat model, we also trust that the driver and kernel communicate via only a set of necessary and predefined interfaces. As a result, even when a function pointer is corrupted, we consider that interface bypass is not viable. 

Statically detecting the second category, interface temporal violations, is generally challenging. While previous works have utilized state machines to describe driver protocols \cite{dingo, formalising_drv_interface, better_device_analysis}, such state machine-based driver development is not commonly employed in existing drivers. Furthermore, existing driver protocols encompass not only the kernel/driver interface but also certain signals from the underlying device. As our static analysis is limited to the kernel/driver interface, capturing all protocol violations statically becomes difficult. To limit our study, we examine interface temporal violations that involve only 
the following protocols: (1) sleep in atomic context; (2) lock and never unlock; and (3) Unbalanced allocation/deallocation. 

\para{SAC violations}
A malicious driver can potentially invoke arbitrary kernel interface functions. For SAC, a sleepable kernel interface must be invoked following a call that obtains a spinlock. If the predefined interface allows the driver to call a spinlock-obtaining function and then a sleepable function, the driver can deliberately violate the expected protocol and trigger SAC. To quantify SAC violation instances, we identify if the target driver can invoke these two types of interface functions, following the method proposed in DSAC~\cite{dsac} to identify sleepable functions. Each unique pair of a spinlock-obtaining function and a sleepable function is considered an instance of an SAC violation.

\para{Lock and never unlock}
A malicious driver can choose to hold a lock by not invoking the corresponding unlock function. To quantify this violation, we analyze the driver code and identify the lock/unlock API functions used by the driver. However, if the obtained lock is a private lock used by the driver, holding it simply causes a DoS attack for the driver, which we do not consider a successful attack on the kernel. Therefore, we first identify shared locks as described in Section~\ref{subsec:quantify_race_condition}, and then identify the lock/unlock calls on these shared locks. Each pair of lock/unlock API functions on a shared lock is considered an \vulName{} instance.

\para{Unbalanced allocation/deallocation}
The number of potential unbalanced allocation/deallocation CIV instances depends on the combination of interface functions that invoke kernel memory allocation and deallocation APIs. Detecting these functions requires a call graph analysis, but the key challenge is determining whether the objects passed to those functions could be the same, which typically requires a scalable and precise alias analysis for large-scale applications like the kernel~\cite{hybrid_alias_analysis_global_var}.
To balance accuracy and scalability, we use a type-based approach: (1) Identify all interface functions that can invoke kernel memory management API; (2) For each pair of allocation and deallocation API calls found, we check if the calls are invoked on objects of the same type. We also consider pointer casts and recovering underlying object types for \lstinline{void*} pointers by tracking type casting operations. For each pair of alloc/dealloc call pairs, if they operate on the same type of object, we consider them as a \vulName{} instance.

%% file: tex/eval.tex
\section{Quantifying CIV Attack Surface} \label{sec:experiment}
\vspace{-0.7em}
In this section, we present our classification results on \vulName{}s. Our evaluation goals are as follows: 
\begin{itemize}
    \item Assuming the baseline threat model, how prevalent is each type of CIV at the driver/kernel isolation boundary across different driver classes? 
    \item How do the baseline CIV statistics change based on the enforcement of extra security properties?
\end{itemize}

To evaluate the effectiveness, we select 11 drivers (Table~\ref{tab:civ_quantification}) from 7 different driver classes for a detailed study. The drivers are selected by referencing previous studies on driver isolation, e.g., Ksplit\cite{ksplit}. These drivers are from major OS subsystems and contain rich interface boundaries, such as the exchange of function pointers in both directions, tagged unions, linked lists, etc.

\subsection{CIV Statistics for the Baseline Threat Model}
\label{subsec:eval_civ_classification}
\vspace{-0.5em}
\para{Shared data leakage} 
Table~\ref{tab:quantify-struct-oversharing} presents the quantification result of data oversharing. The first row shows the number of fields directly accessible to the driver, assuming that all fields are deep-copied to the driver. The second row shows the number of fields needed for correct execution, computed based on the parameter access analysis  by Microdrivers~\cite{microdrivers} (and Decaf~\cite{decaf}). The third row shows the number of kernel/driver shared fields, computed using the shared field analysis by KSplit~\cite{ksplit}. 

The results reveal that the number of shared fields is generally less than 1\% of the number of deep-copied fields and approximately 50\% less than the number of accessed fields. These findings suggest that if objects are shared between the kernel and the driver without any access restrictions, the degree of oversharing is substantial. Even the number of accessed fields (Microdrivers) is still around twice the number of shared fields (KSplit). Based on these observations, we believe that implementing mechanisms to restrict data sharing can significantly reduce the extent for confidentiality-related \vulName{}s stemming from oversharing.

\begin{table*}[htp]
\begin{center}
\vspace{-34pt}
\begin{subtable}[t]{\textwidth}
\resizebox{\textwidth}{!}{
\begin{tabular}{|l|c|c|c|c|c|c|c|c|c|c|c|}
\hline
& \textbf{ixgbe} &  \textbf{null\_net} & \textbf{sfc} &  \textbf{msr} & \textbf{null\_blk} & \textbf{nvme} & \textbf{sb\_edac} & \textbf{mgag200} & \textbf{usb\_f\_fs} & \textbf{power\_meter} & \textbf{dm-zero} \\
\hline
Deep copy & 999K & 48K & 846K & 24K & 227K & 321K & 15K & 467K & 92K &  20K  & 11K \\
\hline
Microdrivers field access \cite{microdrivers} & 4K & 231 & 6K & 66 & 562 & 643 & 91 & 597 & 641 &  144  & 29 \\
\hline
KSplit shared field \cite{ksplit} & 1983 & 73 & 1K & 13 & 247 & 249 & 31 & 333 & 173 & 29  & 46 \\
\hline
\end{tabular}
}
\caption{Quantify overshared struct fields  (data leakage \vulName{}s).}
\label{tab:quantify-struct-oversharing}
\end{subtable}

\vspace{4pt}
\begin{subtable}[t]{\textwidth}
\resizebox{\textwidth}{!}{
\begin{tabular}{|l|*{22}{c|}}
\hline
\vulName{} Classes
& \multicolumn{2}{c|}{\textbf{ixgbe}}
& \multicolumn{2}{c|}{\textbf{null\_net}}
& \multicolumn{2}{c|}{\textbf{sfc}}
& \multicolumn{2}{c|}{\textbf{msr}}
& \multicolumn{2}{c|}{\textbf{null\_blk}}
& \multicolumn{2}{c|}{\textbf{nvme}}
& \multicolumn{2}{c|}{\textbf{sb\_edac}}
& \multicolumn{2}{c|}{\textbf{mgag200}}
& \multicolumn{2}{c|}{\textbf{usb\_f\_fs}}
& \multicolumn{2}{c|}{\textbf{power\_meter}}
& \multicolumn{2}{c|}{\textbf{dm\_zero}} \\
\hline
\hline
\textsc{\textbf{Memory Safety (MEM1)}} & P & M & P & M & P & M & P & M & P & M & P & M & P & M & P & M & P & M & P & M & P & M \\
\hline
MEM1: Pointer value &  48 & 48 & 2 & 1 & 0 & 0 & 0 & 0 & 1 & 1 & 15 & 15 & 2 & 1 & 5 & 3 & 12 & 3 & 0 & 0 & 1 & 1\\
\hline
MEM2: Pointer offset/buffer index & 32 & 24 & 2 & 0 & 0 & 0 & 8 & 1 & 0 & 0 & 10 & 10 & 0 & 0 & 9 & 4 & 9 & 0 & 2 & 2 & 0 & 0\\
\hline
MEM3: Type selector & 0 & 0 & 0 & 0 & 0 & 0 & 0 & 0 & 0 & 0 & 0 & 0 & 0 & 0 & 0 & 0 & 0 & 0 & 3 & 3 & 0 & 0\\
\hline
MEM4: Sensitive kernel memory APIs & 77 & 70 & 5 & 3 & 3 & 3 & 13 & 9 & 2 & 2 & 10 & 10 & 6 & 6 & 26 & 18 & 11 & 9 & 2 & 2 & 2 & 1\\
\hline
MEM5: Corrupted string & 5 & 5 & 2 & 2 & 2 & 2 & 2 & 2 & 4 & 3 & 3 & 2 & 2 & 2 & 3 & 3 & 1 & 1 & 2 & 2 & 0 & 0  \\
\hline
\hline
\textsc{\textbf{Decision-making variable}} & P & M & P & M & P & M & P & M & P & M & P & M & P & M & P & M & P & M & P & M & P & M \\
\hline
DM1: Corrupted guard & 104 & 100 & 45 & 45 & 73 & 73 & 81 & 81 & 20 & 20 & 15 & 13 & 66 & 66 & 72 & 70 & 51 & 51 & 37 & 37 & 2 & 2\\
\hline
DM2: Invalid/wrong error code & 1234 & N/A & 20 & 13 & 1135 & N/A & 9 & 9 & 13 & 10 & 53 & N/A & 1 & 1 & 28 & 28 & 24 & 19 & 4 & 4 & 0 & 0\\ \hline
DM3: Corrupted loop condition & 31 & 25 & 1 & 1 & 2 & 2 & 0 & 0 & 0 & 0 & 10 & 10 & 12 & 9 & 13 & 1 & 7 & 7 & 7 & 7 & 2 & 2\\
\hline
\hline
\textsc{\textbf{Arithmetic error}} & P & M & P & M & P & M & P & M & P & M & P & M & P & M & P & M & P & M & P & M & P & M \\
\hline
AE1: Divided by zero & 0 & 0 & 0 & 0 & 0 & 0 & 0 & 0 & 0 & 0 & 0 & 0 & 0 & 0 & 0 & 0 & 0 & 0 & 0 & 0 & 0 & 0\\
\hline
AE2: Integer overflow/underflow & 28 & 28 & 2 & 0 & 0 & 0 & 9 & 2 & 0 & 0 & 20 & 11 & 0 & 0 & 4 & 2 & 9 & 4 & 0 & 0 & 0 & 0\\
\hline
\end{tabular}
}
\caption{Quantify shared data corruption \vulName{}s. N/A stands for not verified due to large manual effort.}
\label{tab:shared_data_integrity_quantification}
\end{subtable}


\vspace{4pt}
\begin{subtable}[t]{\textwidth}
\resizebox{\textwidth}{!}{
\begin{tabular}{|l|c|c|c|c|c|c|c|c|c|c|c|}
\hline
\multicolumn{1}{|c|}{} & \textbf{ixgbe} & \textbf{null\_net} & \textbf{sfc} & \textbf{msr} & \textbf{null\_blk} & \textbf{nvme} & \textbf{sb\_edac} & \textbf{mgag200} & \textbf{usb\_f\_fs} & \textbf{power\_meter} & \textbf{dm-zero} \\
\hline\hline
No. shared lock & 3 & 1 & 1 & 0 & 0 & 0 & 0 & 0 & 0 & 0 & 0 \\
\hline
\end{tabular}%
}
\caption{\footnotesize Quantify corruptible shared lock (concurrency CIV)}
\label{tab:shared_lock_quantification}
\end{subtable}
\vspace{4pt}
\begin{subtable}[t]{\textwidth}
\resizebox{\textwidth}{!}{
\begin{tabular}{|l|c|c|c|c|c|c|c|c|c|c|c|}
\hline
\multicolumn{1}{|c|}{} & \textbf{ixgbe} & \textbf{null\_net} & \textbf{sfc} & \textbf{msr} & \textbf{null\_blk} & \textbf{nvme} & \textbf{sb\_edac} & \textbf{mgag200} & \textbf{usb\_f\_fs} & \textbf{power\_meter} & \textbf{dm-zero} \\
\hline\hline
Sleep in an atomic context & 3 & 0 & 3 & 0 & 2 & 3 & 0 & 3 & 3 & 0 & 0 \\
\hline
Lock and never unlock & 3 & 1 & 2 & 0 & 4 & 1 & 0 & 1 & 4 & 1 & 0 \\
\hline
Unbalanced alloc/dealloc & 48 & 6 & 64 & 6 & 2 & 47 & 5 & 42 & 14 & 7 & 0 \\
\hline
\end{tabular}%
}
\caption{\footnotesize Quantify control transfer \vulName{}s.}
\label{tab:control_transfer_quantification}
\end{subtable}
\vspace{-10pt}
\caption{\vulName{} quantification results. Table~\ref{tab:quantify-struct-oversharing} shows the degree of overshared fields that are not read by the driver; Table~\ref{tab:shared_data_integrity_quantification} shows the number of shared data integrity CIVs; Table~\ref{tab:shared_lock_quantification} shows the number of corruptible shared lock instances; Table~\ref{tab:control_transfer_quantification} shows the number of control transfer CIVs.}
\label{tab:civ_quantification}
\end{center}
\end{table*}

\para{Shared data corruption} Table~\ref{tab:shared_data_integrity_quantification} presents the statistics of shared data integrity CIVs. Statistics for each subcategory are collected using the taint analysis described in Section~\ref{subsubsec:taint}. Due to the pruning strategy outlined in Section~\ref{subsec:pruning_taint_trace}, we organize the statistics in two columns for each driver: \textit{P} (Pruned) presents the number of paths after applying the pruning strategy, while \textit{M} (Manual) presents the number of CIVs remaining after manual inspection. 

\begin{enumerate}

\item \textbf{Memory Safety:} This subcategory shows the highest diversity of CIVs. MEM1 (pointer value corruption) is particularly prevalent in \lstinline{ixgbe} (48 instances) and \lstinline{nvme} (15 instances). MEM2 (pointer offset/buffer index) also shows significant occurrences in \lstinline{ixgbe} (32 instances). MEM3 (type selector) is rare, with only 3 instances in the \lstinline{power_meter} driver. MEM4 (sensitive kernel APIs) is the most prevalent memory safety \vulName{}, although less common than in CIVScope\cite{CIVScope} due to the consideration of selective data copying in our experiment. MEM5 (corrupted string) is less common, with only 26 instances in total. 

\item \textbf{Decision-making data:} DM1 (corrupted branch guard) instances are found in every driver, indicating that isolated drivers can still affect kernel control paths. DM2 (invalid/wrong error code) is prevalent, with \lstinline{ixgbe} (1,234 instances) and \lstinline{sfc} (1,135 instances) standing out, and notable occurrences in \lstinline{nvme} (53 instances) and \lstinline{mgag200} (28 instances). Some drivers, like \lstinline{ixgbe}, have too many DM2 instances for manual verification. DM3 (Corrupted loop condition) is relatively less common compared to other DM classes (85 instances in total), indicating that many loop conditions within the kernel can potentially be corrupted and leads to buffer overflow when the buffer inside the loop body is accessed.

\item \textbf{Arithmetic Errors:} AE1 (division by zero) is absent across all investigated drivers, indicating that the kernel rarely uses driver-supplied data as a denominator. AE2 (integer overflow/underflow) is more prevalent, with the highest number of vulnerabilities in \lstinline{ixgbe} (28 instances) and \lstinline{nvme} (20 instances), suggesting that arithmetic operations using driver-corrupted data can have potential overflow/underflow issues.
\end{enumerate}

Our static analyses identifies \vulName{}s, which are potential vulnerabilities exist at the interface. However, since static analysis approximates, it cannot confirm that the identified \vulName{}s can cause actual harm to the kernel. To gain confidence on static analysis results, we performed two types of validation: manual validation and construction of proof-of-concept (PoC) exploits for a random sample of cases.


\para{Concurrency} As presented in Table~\ref{tab:shared_lock_quantification}, the \lstinline{ixgbe} driver has the highest instances of corruptible shared lock (3 instances). This is followed by the \lstinline{null_net} and \lstinline{sfc} drivers, each with 1 occurrence. But overall, corruptible shared locks are pretty rare across the studied drivers. This matches KSplit's result\cite{ksplit}, which indicates shared locks are rare across device drivers.

\para{Control transfer} 
Table~\ref{tab:control_transfer_quantification} presents the frequency of interface temporal violation \vulName{}s. SAC vulnerabilities (first row) are relatively infrequent, with \lstinline{ixgbe}, \lstinline{sfc}, \lstinline{nvme}, \lstinline{mgag200}, and \lstinline{usb_f_fs} drivers exhibiting 3 instances each, and \lstinline{null_blk} with 2 instances. Lock and never unlock issues (second row) are slightly more prevalent, with \lstinline{null_blk} and \lstinline{usb_f_fs} having 4 instances each, and \lstinline{ixgbe} having 3 instances. Unbalanced allocation/deallocation issues are prevalent across several drivers, with \lstinline{sfc} leading (64 instances), followed by \lstinline{ixgbe} (48 instances). The fact that almost all drivers have varying degrees of control transfer temporal violations indicates the importance of deploying detection and prevention techniques for such issues, considering that we tested for only a small subset of such temporal violation issues.

\para{Manual validation}
First we conducted a two-phase manual validation on the identified cases. In the first phase, a student evaluated each case to determine if it represented a true positive under our threat model. For each case, the student provides a concise justification, detailing how corrupted shared data could lead to potential attacks. The second phase involved a peer review to confirm the validity of the cases. Our manual validation reveals that the taint analysis achieves a high level of accuracy, with a precision of 90\% across the evaluated cases. This demonstrates the effectiveness of our approach in identifying potential security vulnerabilities. However, the analysis revealed a small number of false positives, primarily due to limitations in our path pruning strategy. Specifically, the second heuristic in our approach relies on data flow analysis to determine whether the data used in taint sinks are subject to proper checks. The imprecision inherent in this heuristic accounts for the observed false positives.



\begin{listing}[t]
\scriptsize
\begin{minted}{c}
void __bitmap_clear(unsigned long *map, unsigned int start, int len) {
  // Vulnerable pointer arithmetic
  unsigned long *p = map + BIT_WORD(start); 
  const unsigned int size = start + len;
  int bits_to_clear = BITS_PER_LONG - (start % BITS_PER_LONG);
  unsigned long mask_to_clear = BITMAP_FIRST_WORD_MASK(start);
  while (len - bits_to_clear >= 0) {
    // Crash with illegal p or a malicious write
    *p &= ~mask_to_clear;
    len -= bits_to_clear;
    bits_to_clear = BITS_PER_LONG;
    mask_to_clear = ~0UL; p++;
  }
}
// manually constructed maliious "driver" code
int main() {
  unsigned long bitmap[2] = {0xFFFFFFFF, 0xFFFFFFFF};
  // Large value to corrupt pointer calculation
  unsigned int large_start = 0x7FFFFFFF;
  // Crash due to invalid pointer arithmetic
  __bitmap_clear(bitmap, large_start, 1);
  return 0;
}
\end{minted}
\caption{An example PoC exploit.}
\label{lst:poc_memory_api}
\end{listing}

\para{Construction of PoC exploits}
To validate the exploitability of the identified \vulName{}s, we randomly sampled a small set of instances from each CIV class, ensuring diversity across different drivers. For each instance, we manually constructed two programs: \textit{K'} (relevant kernel code snippet) and \textit{D'} (attack code for the kernel-driver interface). We then executed these programs to verify if the detected CIVs could be exploited for potential attacks.

For shared data-related CIVs, we instrument the sink operation to check if the malicious values from the driver can reach the sink to control the operation (see Listing~\ref{lst:poc_memory_api} for an example). For control transfer CIVs, we implemented class-specific verification processes. For SAC and Lock and never unlock, we deem the attack successful when the program enters a hang. For unbalanced malloc/dealloc, we consider the attack successful if memory consumption approaches the system's max capacity (128GB) during program execution.


Table~\ref{tab:civ-analysis} summarizes the success rates of PoC exploits for different CIV classes. This process reveals that for a large percentage of instances (60 to 100\%) we could construct successful exploits. Although limited, this study gives evidence that most instances in our static analysis results ( (Table~\ref{tab:civ_quantification}) are real vulnerabilities. During this manual exploit-construction process, we discovered instances where attacks fail, mainly in the shared data corruption category. This is mainly because there are guards after the sink to validate the correctness of sink operations. Due to space limitation, we show a detailed example in Appendix~\ref{app:failed_shared_data_corruption}.

\begin{table}[htp]
\centering
\begin{tabular}{|l|c|}
\hline
\textbf{CIV Class} & \textbf{Success Rate} \\ \hline
\multicolumn{2}{|c|}{\textit{Memory Safety (MEM)}} \\ \hline
MEM1: Pointer value & 9/10 (90\%) \\ \hline
MEM2: Pointer offset/buffer index & 8/10 (80\%) \\ \hline
MEM3: Type selector & 3/3 (100\%) \\ \hline
MEM4: Sensitive kernel memory APIs & 7/10 (70\%) \\ \hline
MEM5: Corrupted string & 8/10 (80\%) \\ \hline
\multicolumn{2}{|c|}{\textit{Decision-making variable (DM)}} \\ \hline
DM1: Corrupted guard & 10/10 (100\%) \\ \hline
DM2: Invalid/wrong error code & 6/10 (60\%) \\ \hline
DM3: Corrupted loop condition & 10/10 (100\%) \\ \hline
\multicolumn{2}{|c|}{\textit{Arithmetic error (AE)}} \\ \hline
AE1: Divided by zero & n/a \\ \hline
AE2: Integer overflow/underflow & 6/10 (60\%) \\ \hline
\multicolumn{2}{|c|}{\textit{Control transfer (CT)}} \\ \hline
CT1: SAC & 10/10 (100\%) \\ \hline
CT2: Lock and never unlock & 10/10 (100\%) \\ \hline
CT3: Unbalanced alloc/free & 10/10 (100\%) \\ \hline
\end{tabular}
\caption{PoC results for CIV classes; ; n/a for "divided by zero" since no such cases were identified during static analysis.}
\label{tab:civ-analysis}
\end{table}

\subsection{Impact of enforcing security properties}
\label{subsec:eval_extra_defenses}
\vspace{-0.5em}
We now analyze the impact of enforcing additional security properties on the prevalence of \vulName{}s. We examine each security property, providing rationale for its potential effect on \vulName{}s. Due to space constraints, the detailed statistics after enforcing these extra security properties are presented in Appendices~\ref{app:civ_reduction_after_advanced_sp}.

\para{Interface data integrity (P3)}
Previous work by Butt et al.\cite{protecting_comm_kernel_from_vulnerable_driver} utilized data invariants to enforce P3, which can be developed manually or through automated methods\cite{daikon}. However, ensuring soundness and completeness of data invariant inference in practice is challenging. While correctly enforcing P3 via data invariants could potentially mitigate all shared data corruption \vulName{}s, the uncertainty of inferable or manually specifiable invariants makes quantifying the impact difficult. Therefore, we do not quantify \vulName{} reduction based on this property.

\para{Control flow integrity (P5)}
By enforcing CFI in a driver, the driver execution is restricted to its control-flow graph (CFG) and thus arbitrary driver code is not possible; attacks can only be performed by data corruption within that CFG (e.g., via control-flow bending~\cite{cfi-bending}).
With CFI enforced, we study whether shared data between the kernel and the driver can be corrupted by memory-unsafe operations in the driver code. We utilize recent work on analyzing memory-safe stack/heap objects~\cite{taming_of_stack,top_of_heap} to measure which shared objects passed to the driver can be accessed through unsafe driver memory operations.  These techniques propose isolating memory-safe objects from unsafe memory accesses (i.e., to memory-unsafe objects) such that we can consider memory-safe objects immune to memory errors, reducing the number of shared data \vulName{}s on such objects (Table~\ref{tab:shared_data_integrity_quantification_reduction}). 


Enforcing CFI can significantly mitigates control-transfer related \vulName{}s as driver execution must follow a CFG. However, CFI would not help if in the CFG there is a path that allows interface temporal violation. For example, for the SAC \vulName{}s, the SAC vulnerabilities that already exist in the driver code will  remain after CFI (Table~\ref{tab:control_transfer_violation_after_protocol_integrity}).

\para{Memory safety (P6)}  
Enforcing memory safety in drivers prevents memory errors from being exploited to corrupt driver memory, reducing shared data corruption CIVs caused by memory errors to zero. However, we note that this does not account for correctness issues in buggy drivers that might still lead to shared data corruption. Memory safety can also mitigate control-transfer related \vulName{}s since is provides stronger protection than CFI. So any control-transfer \vulName{}s that can be mitigated by CFI can also be mitigated by memory safety.



\trent{I disagree with the premise here.  Any bug or malicious driver behavior could be used to exploit the kernel, and there need not be a memory error in the driver.  Could just directly return attack data.}
\gtan{I think one worry is about buggy drivers. Even with mem safety, a buggy driver can still corrupt shared data due to some correctness issue. I tried to integrate Kaiming's writing above, but avoid distinguishing exploitable and malicious drivers, which would bring up a can of worms.} 

\para{Results summary}
The results show the reduction of \vulName{}s after enforcing P5,  presented in Appendix~\ref{app:civ_reduction_after_advanced_sp} due to space limitations. For shared data corruption (Table~\ref{tab:shared_data_integrity_quantification_reduction}), the overall reduction across all classes and drivers ranges from 5\% to 20\%, with an average of approximately 15\%. The Decision Making (DM) class, particularly DM2 (Invalid/wrong error code), shows the most significant improvements, with reductions of up to 186 instances in the ixgbe driver and 170 in the sfc driver. The Memory Safety (MEM) class, especially MEM4 (Sensitive kernel memory APIs), also shows substantial improvements, with reductions of up to 12 instances in the ixgbe driver. Across all evaluated drivers, the ixgbe and sfc drivers shows the most reduction across multiple vulnerability classes.

For control transfer related \vulName{}s (Table~\ref{tab:control_transfer_violation_after_protocol_integrity}), the overall reduction ranges from 83.3\% to 100\% across all categories and drivers. The unbalanced allocation/deallocation class shows the most improvement, with near-complete elimination of CIVs in some drivers. For example, the ixgbe driver sees a reduction from 48 instances to 2 instances (95.8\% reduction), and the sfc driver drops from 64 instances to 2 instances (96.9\% reduction). The SAC and Lock and never unlock categories also demonstrate 100\% reduction where initial vulnerabilities were present. These results indicate that enforcing CFI in drivers (P5) is highly effective in addressing control transfer temporal violations, especially in drivers with initially higher vulnerability counts.

%% file: tex/conclusions.tex
\section{Insights for Driver Isolation Frameworks}
\label{sec:lessons_future_iso}
\vspace{-0.4em}
Based on our results, we have identified several key insights that can be used to improve the security of future driver isolation projects. 

\para{L1: Importance of modeling security properties}
Recent driver isolation techniques have made significant strides, featuring low-overhead isolation with advanced hardware support~\cite{lvds, sfitag} and lightweight deployment through automation~\cite{ksplit}. However, most of these approaches still assume a buggy driver threat model, leading to the enforcement of only basic security properties, such as P1 and P4. Our analysis reveals that when considering an exploitable driver threat model, there are plenty of \vulName{}s can be exploited by the driver to attack the kernel through the isolation interfaces. This implies the need to explicitly define both the assumed driver threat model and the expected security properties, which is essential for comprehensively understanding the isolation guarantees and enables a more accurate assessment of the security benefits offered by a given isolation framework.

\para{L2: Securing interfaces is feasible}
In our experiment, we observed that the number of cases for CIVs was significantly lower than those reported in previous studies that employed static analysis for CIV identification~\cite{CIVScope}. This reduction can be attributed to our implementation of a set of pruning strategies and our focus on shared data. We believe these lower numbers suggest that effective defense against CIVs is possible without imposing drastic changes to the interface.

\para{L3: Selective data sharing for complex kernel objects can reduce data leakage \vulName{}s}
The severity of shared data leakage issues can be escalated based on the degree of oversharing, particularly when it involves large and complex kernel objects, such as structured objects. To mitigate this risk, it is beneficial to explore the viability of deploying some form of shared data access control. Our analysis demonstrates that implementing access control mechanisms can significantly reduce the amount of overshared data, thereby helping to enforce kernel data confidentiality. 

\para{L4: Enforcing CFI and memory safety in the driver helps hardening the isolation interface}
In addition to selective data sharing, our analysis highlights the importance of enforcing security properties directly within the driver itself. By enforcing properties such as P5 and P6, the isolation framework can effectively reduce the attack surface for \vulName{}s. This indicates that, besides the driver isolation, future isolation efforts can also gain security benefits by using existing kernel hardening techniques, such as CFI, or develop drivers in safe programming language such as Rust.


\section{Conclusion} \label{sec:conclusion}
\vspace{-0.5em}
We have systemized existing driver isolation frameworks to understand their enforced security properties. We also investigated existing taxonomy of \vulName{}s, and estimate their impact under a baseline threat model. In addition, we explore how enforcing security properties can mitigate \vulName{}s. We believe that future development on driver isolation frameworks can benefit from our findings.

%% file: tex/acks.tex

%% file: tex/appendix.tex
\appendix
\subsection{Unbalanced allocation and free example}\label{app:unbalanced_allocation_free}
\begin{listing}[thp]
\begin{minted}[linenos,
               fontsize=\scriptsize]{c}
int bond_create(struct net *net, const char *name)
{
    struct net_device *bond_dev;
    struct bonding *bond;
    int res = -ENOMEM;

    rtnl_lock();  
    
    bond_dev = alloc_netdev_mq(sizeof(struct bonding), ...);  
    
    if (!bond_dev)  
        goto out;  
    
    bond = netdev_priv(bond_dev);  
    dev_net_set(bond_dev, net);  
    bond_dev->rtnl_link_ops = &bond_link_ops;  
    
    res = register_netdevice(bond_dev);  
    if (res < 0) {  
        free_netdev(bond_dev);  
        goto out;  
    }  
    ... 
    out:
    rtnl_unlock();
    return res;
}
\end{minted}
\caption{Vulnerable Linux ethernet bonding driver with unbalanced memory allocation.}
\label{lst:vulnerable-bonding-driver}
\end{listing}

List~\ref{lst:vulnerable-bonding-driver} shows part of the code of the Linux ethernet bonding driver. In the \lstinline{bond_create} function, memory is allocated for the bonding network device using \lstinline{alloc_netdev_mq} (Line 7). If the registration is not successful, the allocated object is freed and in the error-handling path. However, a malicious driver, with the ability to execute arbitrary code, can deliberately skip the error handling, causing the unregistered object to persist in memory. This results in a memory leak, as the allocated memory for the network device remains occupied but unreachable. Over time, repeated allocation by the malicious driver can accumulate significant amounts of leaked memory, potentially exhausting system resources.

\subsection{Failed shared data corruption example}\label{app:failed_shared_data_corruption}

\begin{listing}[h]
\begin{minted}{c}
static void *kmalloc_reserve(unsigned int *size, gfp_t flags, ...) {
    ... 
    size_t obj_size = SKB_HEAD_ALIGN(*size);
    obj = kmalloc_node_track_caller(obj_size, flags, ...);
    if (obj || !(gfp_pfmemalloc_allowed(flags)))
        goto out;
    ...
    out:
    ...
}
\end{minted}
\caption{Example of check comes after potentially corrupted sink for shared data corruption.}
\label{lst:check_after_sink}
\end{listing}

During the manual verification process, we discovered instances where guards are implemented after the sink to validate the correctness of sink operations, as illustrated in Listing~\ref{lst:check_after_sink}. In the example, both the \lstinline{obj_size} and flags parameters to the \lstinline{kmalloc_node_track_caller} function could be corrupted by the driver. However, after the callsite, the allocated object and flags are both checked. In these cases, we classify such instances as benign and mark them as non-exploitable in our verification process.

\subsection{Shared data corruption/Control transfer \vulName{} number after enforcing CFG (P5)}
\label{app:civ_reduction_after_advanced_sp}
\begin{table*}[h]
\resizebox{\textwidth}{!}{
\begin{tabular}{|l|*{11}{c|}}
\hline
\vulName{} Classes
& \textbf{ixgbe}
& \textbf{null\_net}
& \textbf{sfc}
& \textbf{msr}
& \textbf{null\_blk}
& \textbf{nvme}
& \textbf{sb\_edac}
& \textbf{mgag200}
& \textbf{usb\_f\_fs}
& \textbf{power\_meter}
& \textbf{dm\_zero} \\
\hline
\hline
\textsc{\textbf{Memory Safety (MEM1)}} & & & & & & & & & & & \\
\hline
MEM1: Pointer value & 41 & 2 & 0 & 0 & 1 & 13 & 2 & 4 & 10 & 0 & 1 \\
\hline
MEM2: Pointer offset/buffer index & 27 & 2 & 0 & 7 & 0 & 8 & 0 & 8 & 8 & 2 & 0 \\
\hline
MEM3: Type selector & 0 & 0 & 0 & 0 & 0 & 0 & 0 & 0 & 0 & 3 & 0 \\
\hline
MEM4: Sensitive kernel memory APIs & 65 & 4 & 3 & 11 & 2 & 9 & 5 & 22 & 9 & 2 & 2 \\
\hline
MEM5: Corrupted string & 4 & 2 & 2 & 2 & 3 & 3 & 2 & 3 & 1 & 2 & 0 \\
\hline
\hline
\textsc{\textbf{Decision-making variable}} & & & & & & & & & & & \\
\hline
DM1: Corrupted guard & 88 & 38 & 62 & 69 & 17 & 13 & 56 & 61 & 43 & 31 & 2 \\
\hline
DM2: Invalid/wrong error code & 1048 & 17 & 965 & 8 & 11 & 45 & 1 & 24 & 20 & 3 & 0 \\ 
\hline
DM3: Corrupted loop condition & 26 & 1 & 2 & 0 & 0 & 8 & 10 & 11 & 6 & 6 & 2 \\
\hline
\hline
\textsc{\textbf{Arithmetic error}} & & & & & & & & & & & \\
\hline
AE1: Divided by zero & 0 & 0 & 0 & 0 & 0 & 0 & 0 & 0 & 0 & 0 & 0 \\
\hline
AE2: Integer overflow/underflow & 24 & 2 & 0 & 8 & 0 & 17 & 0 & 3 & 8 & 0 & 0 \\
\hline
\end{tabular}
}
\caption{Quantify shared data corruption \vulName{}.}
\label{tab:shared_data_integrity_quantification_reduction}
\end{table*}
\begin{table*}[h]
\resizebox{\textwidth}{!}{
\begin{tabular}{|l|c|c|c|c|c|c|c|c|c|c|c|}
\hline
\multicolumn{1}{|c|}{} & \textbf{ixgbe} & \textbf{null\_net} & \textbf{sfc} & \textbf{msr} & \textbf{null\_blk} & \textbf{nvme} & \textbf{sb\_edac} & \textbf{mgag200} & \textbf{usb\_f\_fs} & \textbf{power\_meter} & \textbf{dm-zero} \\
\hline\hline
SAC & 0 & 0 & 0 & 0 & 0 & 0 & 0 & 0 & 0 & 0 & 0 \\
\hline
Lock and not unlock & 0 & 0 & 0 & 0 & 0 & 0 & 0 & 0 & 0 & 0 & 0 \\
\hline
Unbalanced allocation/deallocation & 2 & 1 & 2 & 0 & 0 & 0 & 0 & 0 & 0 & 0 & 0 \\
\hline
\end{tabular}%
}
\caption{Control transfer \vulName{}s after enforcing CFI (P5).}
\label{tab:control_transfer_violation_after_protocol_integrity}
\end{table*}